\documentclass[12pt]{iopart}
\eqnobysec
\usepackage{iopams} 
\usepackage{epsfig}
\usepackage{amssymb}
\begin{document}

\title[Loss of non-Gaussianity for damped photon-subtracted thermal states] {Loss of non-Gaussianity for damped photon-subtracted thermal states}
\author{
Iulia Ghiu{$^1$},
Paulina Marian{$^{1,2}$},
and Tudor A. Marian{$^1$}}

\address{$^1$Centre for Advanced  Quantum Physics,
Department of Physics, University of Bucharest, 
R-077125 Bucharest-M\u{a}gurele, Romania}

\address{$^2$ Department of Physical Chemistry,
University of Bucharest, Boulevard Regina Elisabeta 4-12, R-030018  Bucharest, Romania}
\ead{iulia.ghiu@g.unibuc.ro\\
paulina.marian@g.unibuc.ro\\
tudor.marian@g.unibuc.ro}

\begin{abstract} We investigate non-Gaussianity properties for a set 
of classical one-mode states obtained by subtracting  photons from a thermal state. Three distance-type degrees of non-Gaussianity  used for these states are shown to have a monotonic behaviour
with respect to their mean photon number. Decaying of their non-Gaussianity under damping is found to be consistently described by the distance-type measures considered here. We also compare the dissipative evolution of non-Gaussianity when starting from 
$M$-photon-subtracted and $M$-photon-added thermal states. 

 \end{abstract}


\maketitle

\section{Introduction}

In quantum optics, non-Gaussian states were studied in connection
with some non-classical properties such as photon antibunching and quadrature or amplitude-squared squeezing. A survey on 
non-classicality defined as the non-existence of the Glauber-Sudarshan $P$ representation as a genuine probability density can be found in Ref. \cite{1}. Interest in the non-Gaussian states  has then renewed
in quantum information processing due to their efficiency in some quantum  protocols \cite{Dakna1,Opat}. In general, their usefulness was connected to certain non-classicality properties detected by  negative values of the Wigner function. However, it has recently been realized that quantum states with non-negative (Gaussian or 
non-Gaussian) Wigner functions can be efficiently simulated on a classical computer \cite{ME,VWFE}. It appears that non-Gaussianity of a quantum state is a useful feature in quantum information processing, regardless of being non-classical or classical according to the concepts of quantum optics. To quantify this property as a resource,  some non-Gaussianity measures were recently defined as  distances between the given state $\hat \rho$ and its associate Gaussian state 
$\hat \tau_G$. Here $\hat \tau_G$ is the unique Gaussian state having the same mean displacement and covariance matrix as $\hat \rho$ \cite{P2,P3,P33,GMM,MGM}. In Refs. \cite{P2,P3,P33}, Genoni {\em et al.} 
used the Hilbert-Schmidt metric and the relative entropy as distances and gave a comprehensive discussion of the general properties of non-Gaussianity degrees for large sets of one-mode, two-mode, and multimode states.  Later, in Refs. \cite{GMM,MGM} a degree of non-Gaussianity based on the Bures metric was similarly introduced. 
We stress now that all the distance-type measures considered so far 
used the same state $\hat \tau_{\rm G}$ as a reference Gaussian state in evaluating non-Gaussianity. It was only very recently that two 
of us succeeded to prove that the relative entropy of {\em any} 
$N$-mode state to its associate Gaussian one $\hat \tau_{\rm G}$ is 
an {\em exact} distance-type measure of non-Gaussianity \cite{PT2013}. 

All these measures were already employed to evaluate non-Gaussianity degrees in some interesting experiments. In the experiment reported in Ref. \cite{B} the relative-entropy measure was used for single-photon-added coherent states, while in Ref. \cite{Ol} the same degree was evaluated in an experiment with multiple-photon subtraction from a thermal state.  The three above-mentioned degrees of non-Gaussianity were determined and compared in some recent experiments 
on phase-averaged coherent states \cite{Ol1,Ol2}. 

The present work parallels some of our recent findings on the 
non-Gaussianity and its decay in contact with a thermal reservoir 
for an interesting class of Fock-diagonal one-mode states: 
the photon-added thermal states \cite{GMM,MGM}. Here we intend to compare the three distance-type amounts of non-Gaussianity 
during the damping of two different excitations on a single-mode thermal state: an $M$-photon-added thermal state (PATS) and 
an $M$-photon-subtracted thermal state (PSTS). 

The plan of our paper is as follows. In Section 2 we recall some statistical properties of a PSTS. In Section 3 the three usual 
degrees of non-Gaussianity for a Fock-diagonal one-mode state are recapitulated. Then we derive an analytic expression 
of the Hilbert-Schmidt measure of non-Gaussianity for a PSTS. 
Plots of the entropic and Bures amounts of non-Gaussianity are shown to be in agreement with the Hilbert-Schmidt measure. 
Section 4 examines the  evolution of a PSTS due to the interaction
of the field mode with a thermal reservoir, which is governed 
by the quantum optical master equation. We finally compare
the decay of non-Gaussianity for pairs of states,  PSTSs and PATSs,
having the same thermal mean occupancy $\bar n$, as well as the same number $M$ of subtracted and, respectively, added  photons.
Our concluding remarks are presented in Section 5.

\section{Multiple-photon-subtracted thermal states}

It is known that excitations on a Gaussian state $\hat \rho_{G}$ 
of the type $\hat\rho \sim (\hat a^{\dag})^k \;\hat a^l\;
\hat \rho_{G}\; (\hat a^{\dag})^l \;\hat a^k $ lead to non-Gaussian states \cite{AT,ZF,DKMM,Dakna1,MD}.  Here $\hat a$ and $\hat a^{\dag}$ are the amplitude operators of the field mode. Interesting experiments 
on such states were recently conducted to enlighten their fundamental features \cite{WTG,OTLG,PZK,ZPKB,Ol}. Addition of photons to any classical Gaussian state, in particular to a coherent or a thermal one, generates a non-Gaussian output which is no longer classical 
\cite{AT}. In Refs. \cite{GMM,MGM} we have recently studied 
the non-Gaussianity of $M$-photon-added thermal states and investigated their behaviour under damping. Their non-classicality 
was marked by the negativity of both Wigner and $P$ functions surviving to some extent under damping as well. On the contrary, 
an $M$-photon-subtracted Gaussian state can be either classical 
or non-classical, depending on the input state $\hat \rho_{G}$ 
\cite{KPKJ,BA,WHF}. For instance, subtraction of photons from 
a coherent or a thermal state provides a classical non-Gaussian output \cite{KPKJ,K}. Therefore, the PSTSs are an interesting example 
of {\em classical} non-Gaussian Fock-diagonal states. Measurement 
of their photon statistics by means of photon-number-resolved detection was recently reported \cite{Zhai}. In this section 
we recall some of their statistical properties. 

We thus consider an arbitrary single-mode PSTS:
\begin{equation}
\hat{\rho}_M^{\, \rm sub}(\bar n)=\frac{1}{M!\, (\bar n)^M}\,\hat a^M \, \hat \rho_{\, \rm T}(\bar n)\, (\hat a^\dagger )^M,
\qquad (M=1, 2, 3, \dots).  
\label{PSTS}
\end{equation}
Here $M$ is the number of photons extracted from the mode
and $\hat \rho_{\, \rm T}(\bar n)$ is a thermal state whose mean 
number of photons is  $\bar n>0$:
\begin{equation}
\hat \rho_{\,\rm T}(\bar n)=(1-x)\sum_{n=0}^{\infty}\, x^n \, |n\rangle \langle n| \quad {\rm with} \quad x:=\frac{\bar n}{\bar{n}+1}>0. 
\label{TS}
\end{equation}
A PSTS\ (\ref{PSTS}) is Fock-diagonal, with the photon-number
probabilities 
\begin{eqnarray}
p_n^{\, \rm sub}:=[{\rho}_M^{\, \rm sub}(\bar n)]_{nn}=
\left( \begin{array}{c} 
n+M\\ M 
\end{array} \right) (1-x)^{M+1}x^n, \;\; (n=0,1,2,3, \dots). 
\label{pnd}
\end{eqnarray}
This is actually a negative binomial distribution \cite{Feller}
with the stopping parameter $r=M+1$. Its generating function is
\begin{equation}
{\cal G}_M^{\, \rm sub}(\bar n, v):=\sum_{n=0}^{\infty} p_n^{\, \rm sub}\, v^n =\left( \frac{1-x}{1-xv}\right )^{M+1}, \qquad 
(-1 \leqq v \leqq 1).
\label{genf}
\end{equation}
Accordingly, the mean number of photons in the PSTS\ (\ref{PSTS}),
\begin{equation}
\langle \hat n \rangle:=\sum_{n=0}^{\infty} n\,  p_n^{\, \rm sub}
=\left[ \frac{\partial}{\partial v}\, {\cal G}_M^{\, \rm sub}
(\bar n, v)\right] _{v=1},
\label{occup}
\end{equation}
has the expression:
\begin{equation}
\langle \hat n \rangle=(M+1)\bar n.
\label{mnp}
\end{equation}
It is therefore proportional to the thermal mean occupancy $\bar n$ and, rather counterintuitively, increases with the number $M$ of extracted photons.
Note that the  purity of a PSTS\ (\ref{PSTS}),
\begin{equation}
{\rm Tr}\{[\hat{\rho}_M^{\, \rm sub}(\bar n)]^2 \}
=\sum_{n=0}^{\infty}\, \left( p_n^{\, \rm sub}\right) ^2
=(1-x)^{2(M+1)}{_{2}F_{1}}(M+1, M+1; 1 ; x^2),
\label{pur}
\end{equation}
where ${_{2}F_{1}}$ is a Gauss hypergeometric function\ (\ref{a1}),
coincides with that of the PATS $\hat{\rho}_M^{\, \rm add}(\bar n)$, which was written in Ref. \cite{GMM}:
\begin{equation}
{\rm Tr}\{[\hat{\rho}_M^{\, \rm sub}(\bar n)]^2 \}
={\rm Tr}\{[\hat{\rho}_M^{\, \rm add}(\bar n)]^2 \},
\qquad (M=1, 2, 3, \dots). 
\label{pura}
\end{equation}
An equivalent form of Eq.\ (\ref{pur}) in terms of a Legendre polynomial ${\cal P}_M$ can be obtained by using Eqs.\ (\ref{a2}) 
and\ (\ref{a3}): 
\begin{equation}
{\rm Tr}\{[\hat{\rho}_M^{\, \rm sub}(\bar n)]^2 \}
={\left(  \frac{1-x}{1+x}\right) }^{M+1}{\cal P}_M\left( 1
+ \frac{2x^2}{1-x^2}\right).
\label{purLeg}
\end{equation}

\section{Non-Gaussianity of a photon-subtracted thermal state}
 
We do not insist on the general properties of the distance-type degrees of non-Gaussianity introduced in Refs. \cite{P2,P3,GMM}. 
We just recall the three degrees of non-Gaussianity written 
for a Fock-diagonal state $\hat \rho$ (our case in the following),
\begin{eqnarray} 
\hat \rho=\sum_{n=0}^{\infty} p_n \, |n\rangle \langle n| \quad
{\rm with} \quad \sum_{n=0}^{\infty}  p_n=1. 
\label{diag0} 
\end{eqnarray}
In this case, the associate Gaussian state is a thermal one with the same mean photon occupancy, 
$\langle \hat n \rangle=\sum_{n=0}^{\infty}n\,p_n$:
\begin{eqnarray} 
\hat\tau_{\rm G}=\sum_{n=0}^{\infty} s_n |n\rangle \langle n| \quad{\rm with} \quad
s_n:=\frac{1}{\langle \hat n \rangle+1}\, \sigma^n, \quad 
\sigma:=\frac{\langle \hat n \rangle}{\langle \hat n \rangle+1}.
\label{diag1}
\end{eqnarray}
The Hilbert-Schmidt and entropic amounts of non-Gaussianity were written in Refs. \cite{P2,P3,GMM} as:
\begin{eqnarray}
\delta_{\rm HS}[\hat \rho]&=&\frac{1}{2}\left[ 1+\frac{\sum_{n}(s_n^2
-2 s_n\,p_n)}{\sum_{n}p_n^2}\right]
\nonumber\\  &&=\frac{1}{2}
+\frac{1}{2\, {\rm Tr}\left( {\hat \rho}^2 \right)}\left[ \frac{1}{2
\langle \hat n \rangle+1}-\frac{2}{\langle \hat n \rangle+1}\,
{\cal G}_{\hat \rho}(\sigma) \right], 
\label{HS2}
\end{eqnarray}
and, respectively,
\begin{eqnarray} 
\delta_{\rm RE}[\hat \rho]=\sum_{n=0}^{\infty}\, p_n\ln{p_n}+
(\langle \hat n\rangle+1)\ln (\langle \hat n \rangle+1)-\langle \hat n \rangle\ln(\langle \hat n \rangle).
\label{re2}
\end{eqnarray}
Here we have used the purity of the thermal state $\hat \tau_G$ arising from Eq.\ (\ref{diag1}), while  
${\cal G}_{\hat \rho}(y):=\sum_{n=0}^{\infty} p_n\, y^n$ is the generating function of the photon-number distribution in the given state $\hat \rho$.

The third measure of interest was introduced in terms of the Bures distance between the state  $\hat \rho$ and its associate Gaussian state ${\hat \tau}_{\rm G}$  \cite{GMM,MGM}. In the case 
of a Fock-diagonal state,  we notice the commutation relation 
$[\hat \rho,\hat \tau_{\rm G}]=\hat 0,$  which implies the simpler formula:
\begin{eqnarray}
\delta_{\rm F}[\hat\rho]=1-\sum_{n=0}^{\infty} \sqrt{ p_n\, s_n}.
\label{f}
\end{eqnarray}

The Hilbert-Schmidt degree of non-Gaussianity, Eq.\ (\ref{HS2}), 
can readily be evaluated for any PSTS. First, we employ 
Eqs.\ (\ref{mnp}) and\ (\ref{diag1}) to write the generating function \ (\ref{genf}) for $v=\sigma$: 
\begin{equation}
{\cal G}_M^{\, \rm sub}(\bar n, \sigma)=\left[\frac{(M+1)\bar n +1}
{(M+2)\bar n +1}\right]^{M+1}.
\label{genf1}
\end{equation}
Then, by replacing Eqs.\ (\ref{purLeg}) and\ (\ref{genf1}) into
Eq.\ (\ref{HS2}), one finds the formula: 
\begin{eqnarray}
\delta_{\rm HS}\left[ \hat{\rho}_M^{\, \rm sub}(\bar n)\right]
&=&\frac{1}{2} +\frac{(2\bar n +1)^M}{2\, {\cal P}_M \left( 1
+ \frac{2{\bar n}^2}{2\bar n +1}\right)} 
\left\{ \frac{1}{1+\frac{2M\, \bar n}{2\bar n +1}} \right.
\nonumber\\  && \left.
-\frac{2}{1+\frac{M\, \bar n}{2\bar n +1}}
\left[\frac{(M+1)\bar n +1}{(M+2)\bar n +1}\right]^M \right\}. 
\label{HS3}
\end{eqnarray}
Plots of the Hilbert-Schmidt measure\ (\ref{HS3}) versus the number 
$M$ of subtracted photons at some values of the thermal mean occupancies are shown in Figure 1 (right). In Figure 2 (right) we keep constant the value of $M$ and give the dependence of $\delta_{\rm HS}$ on the thermal parameter $x$.
Also plotted in Figures 1 and 2 are the entropic and Bures degrees 
of non-Gaussianity, Eqs.\ (\ref{re2}) and\ (\ref{f}), respectively. Here we have performed numerically the corresponding summations 
making use of the probability distributions\ (\ref{pnd}) 
and\ (\ref{diag1}).
\begin{figure*}[h]
\center
\includegraphics[width=5cm]{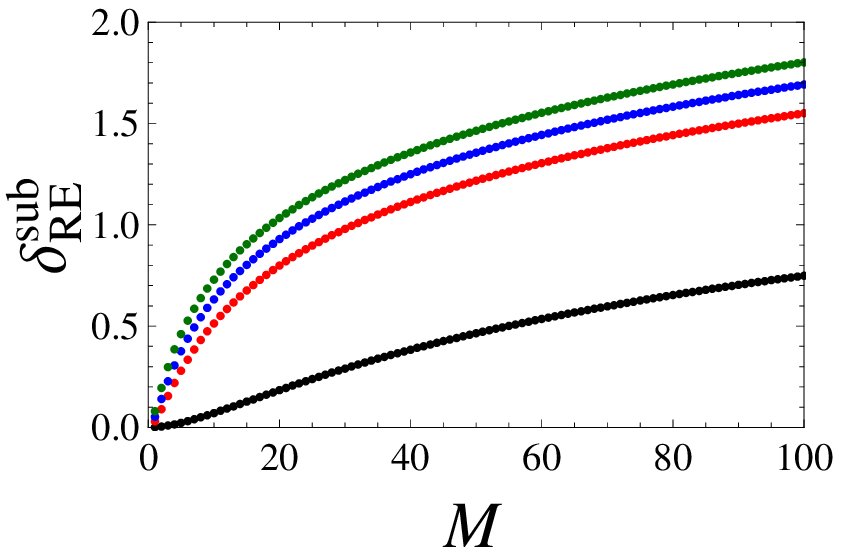}
\includegraphics[width=5cm]{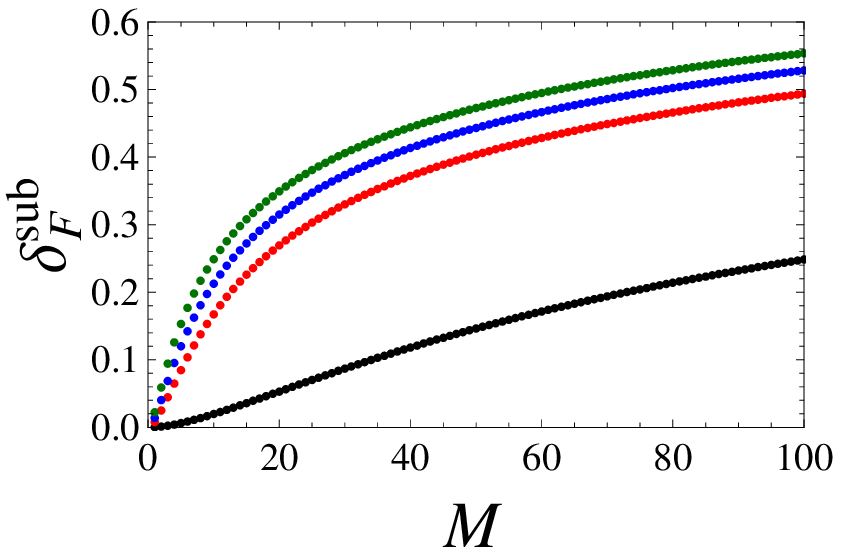}
\includegraphics[width=5cm]{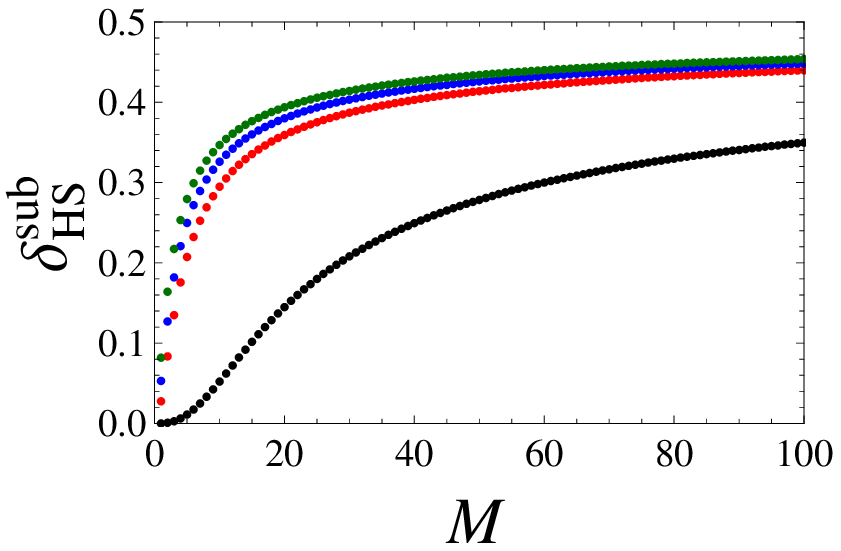}
\caption{Dependence of the distance-type measures of non-Gaussianity on the number of subtracted photons. All the plots start from origin. The lowest curve is for $\bar n=0.1$. For the upper ones we have used 
$\bar n=1, 2, 5$, respectively. }
\label{fig-1}
\end{figure*}
As in the case of a PATS \cite{GMM}, non-Gaussianity of a PSTS increases with the number of subtracted photons. As one can see 
from Figure 1 of Ref. \cite{GMM}, when adding photons to a thermal state, non-Gaussianity decreases with the thermal parameter $x$, 
which is not the case for a PSTS in our present Figure 2.
\begin{figure*}[h]
\center
\includegraphics[width=5cm]{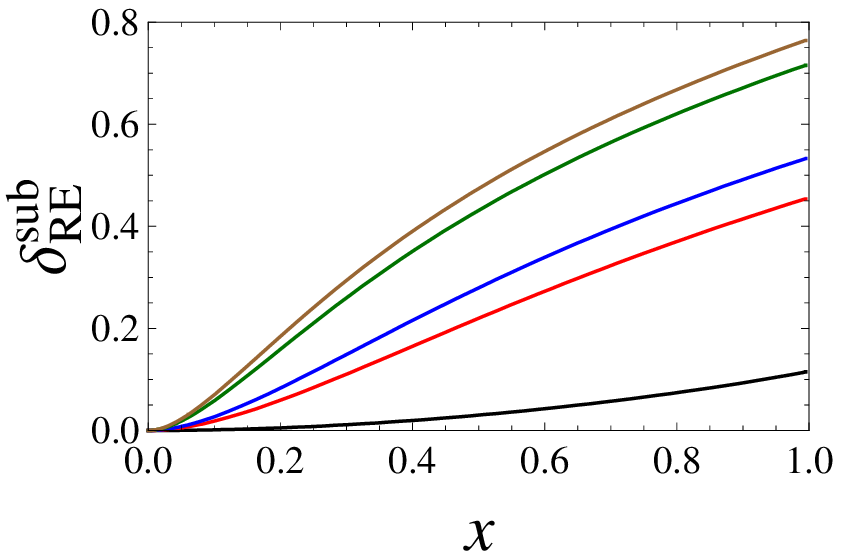}
\includegraphics[width=5cm]{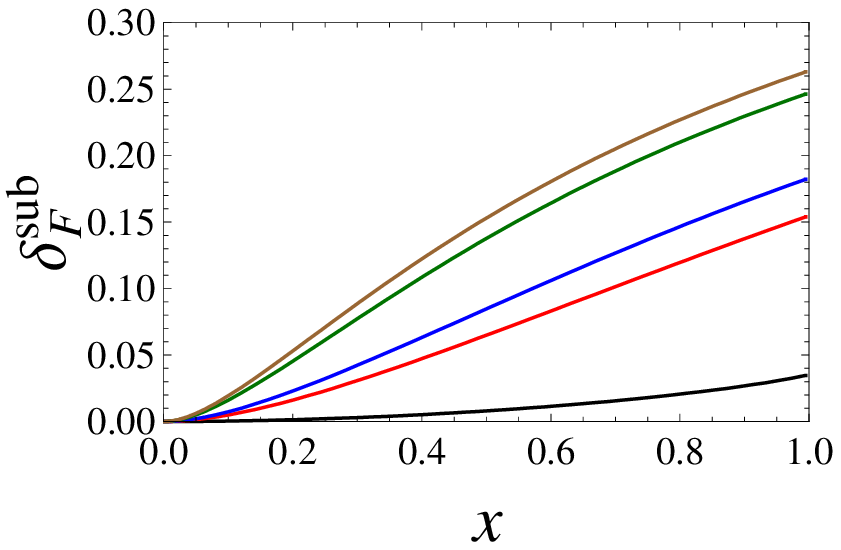}
\includegraphics[width=5cm]{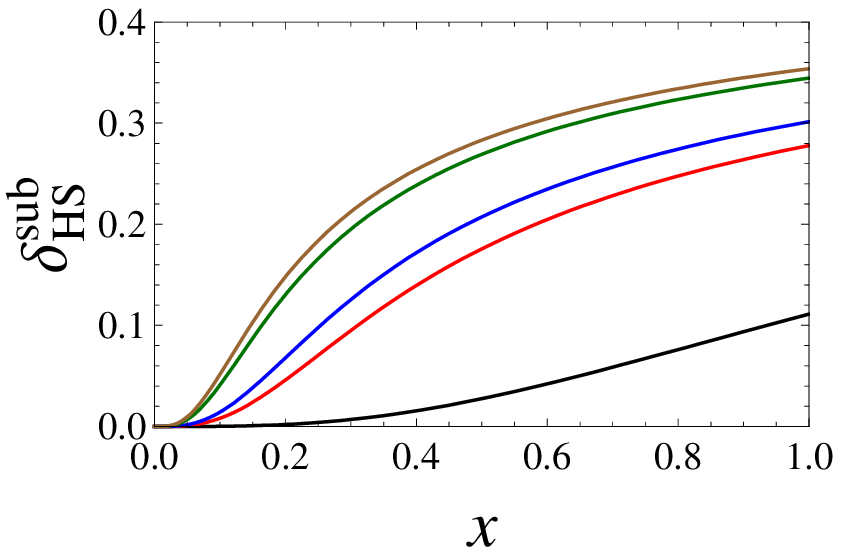}
\caption{Dependence of the distance-type measures of non-Gaussianity on the thermal parameter $x$ for $M$-photon-subtracted thermal states. All the plots start from origin. The lowest plot is for $M=1$. 
For the upper ones we have used $M=4,5,8,9$, respectively. }
\label{fig-2}
\end{figure*}
We have to remark that the three measures we used here give similar dependencies on the parameters involved, which is a signature 
of their consistency. On physical grounds, we expect that a good measure of non-Gaussianity has a monotonic behaviour with respect 
to the mean photon number $\langle \hat n\rangle$ and, in turn, 
to the parameters entering its expression\ (\ref{mnp}). According to our plots in Figures 1 and 2, this is verified for the class of PSTSs for all the measures investigated here.

\section{Gaussification by damping}

In our paper \cite{MGM} the evolution under the quantum optical 
master equation of a Fock-diagonal density matrix  was conveniently written in the interaction picture: 
\begin{eqnarray}
& \rho_{jk}(t)=\delta_{jk}\frac{[{\bar n}_{T}(t)]^j}
{[{\bar n}_{T}(t)+1]^{j+1}}\, \sum_{l=0}^{\infty}\rho_{ll}(0)
\left[\frac{({\bar n}_R+1)
(1-e^{-\gamma t})}{{\bar n}_{T}(t)+1}\right]^l
\nonumber\\ 
& \times {_2F_1}\left[ -j, -l;\,1\,;\frac{e^{-\gamma t}}
{({\bar n}_R+1)(1-e^{-\gamma t}){\bar n}_{T}(t)}\right].
\label{dm1}
\end{eqnarray} 
In Eq.\ (\ref{dm1}), ${\bar n}_R$ and $\gamma$ are constants of the thermal bath
and ${\bar n}_{T}(t):={\bar n}_R (1-e^{-\gamma t})$. The limit 
$t\rightarrow \infty$ in Eq.\ (\ref{dm1}) represents a thermal state with the Bose-Einstein mean photon occupancy $\bar{n}_R$. We thus 
deal with an evolving {\em Gaussification} process which eventually destroys both the non-Gaussianity and the non-classicality properties of any input state. The corresponding time-dependent associate  Gaussian state is a thermal one whose mean occupancy is equal to the average photon number of the damped field state. We find:
\begin{equation}
\langle \hat n \rangle |_t=\left[\bar n (M+1)\right] 
e^{-\gamma t}+{\bar n}_{T}(t).
\label{num1}
\end{equation}
By employing Eqs.\ (\ref{a2}) and\ (\ref{a4}) we obtain the following expression of the photon-number distribution in a damped PSTS:
\begin{eqnarray}
&& p_n^{\, \rm sub}(t)={\left[\hat{\rho}_M^{\, \rm sub}(\bar n)
\right]}_{nn} \big|_t=\frac{[{\bar n}_T(t)+1]^M\, 
[\bar n{\rm e}^{-\gamma t}+{\bar n}_T(t)]^n} 
{{[\bar n{\rm e}^{-\gamma t}+\bar n_T(t)+1]}^{M+n+1}} 
\nonumber \\
&&\times  {_2F_1}\left(-M, -n;1;\frac{\bar n\, {\rm e}^{-\gamma t}}{[{\bar n}_T(t)+1][\bar n{\rm e}^{-\gamma t}+{\bar n}_T(t)]}\right).
\label{pndd}
\end{eqnarray}
We have used the time-dependent probability distribution\ (\ref{pndd}) 
and the mean photon occupancy\ (\ref{num1}) to evaluate numerically 
two of the distance-type measures of non-Gaussianity we are interested in: $\delta_{\rm RE}[\hat \rho_{M}^{\, \rm sub}(\bar n)]|_t$ and  
$\delta_{\rm F}[\hat \rho_{M}^{\, \rm sub}(\bar n)]|_t$  
via Eqs.\ (\ref{re2}) and\ (\ref{f}), respectively. The last degree 
of non-Gaussianity we consider here is the Hilbert-Schmidt one, 
Eq.\ (\ref{HS2}). As in the case of $M$-photon added thermal states \cite{MGM}, the necessary ingredients to evaluate the Hilbert-Schmidt degree of non-Gaussianity can be obtained analytically. However, 
we do not write here the explicit expressions of the time-dependent purity and generating function. Our results are displayed in 
Figure 3, where the time evolutions of the three non-Gaussianity measures are presented for the same values of the parameters.
\begin{figure*}[h]
\center
\includegraphics[width=5cm]{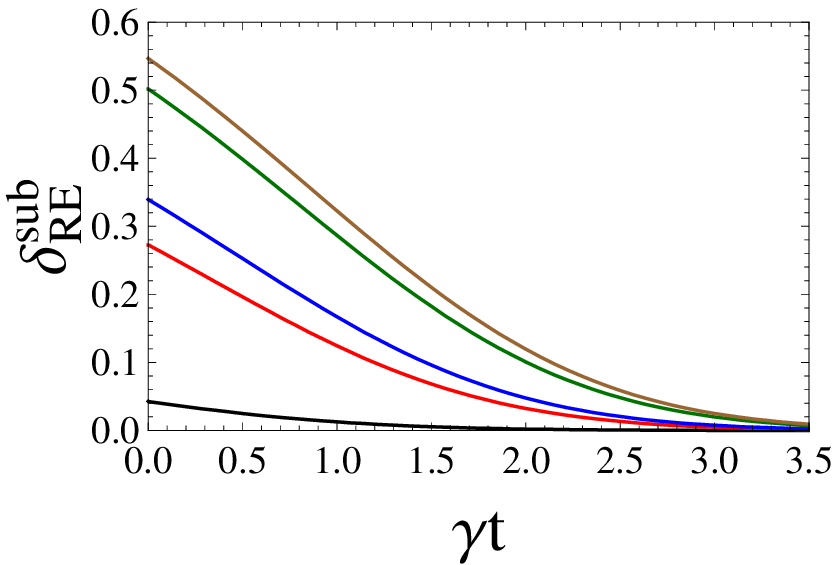}
\includegraphics[width=5cm]{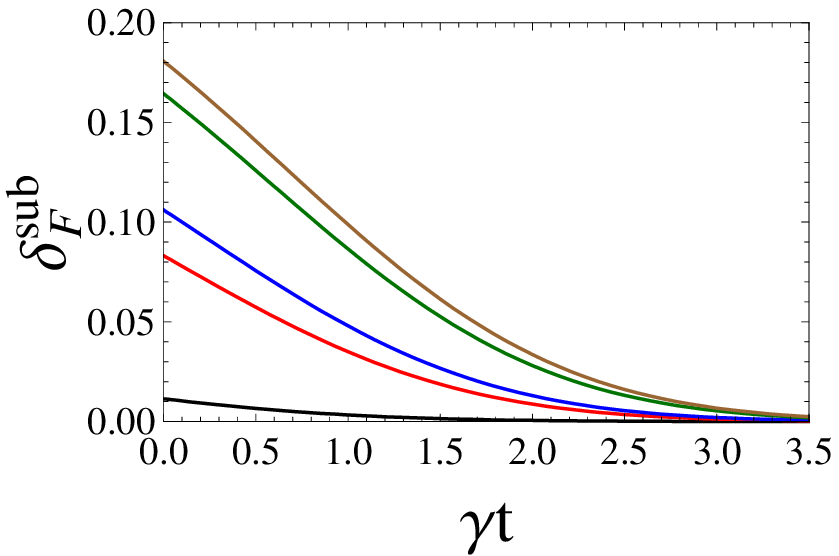}
\includegraphics[width=5cm]{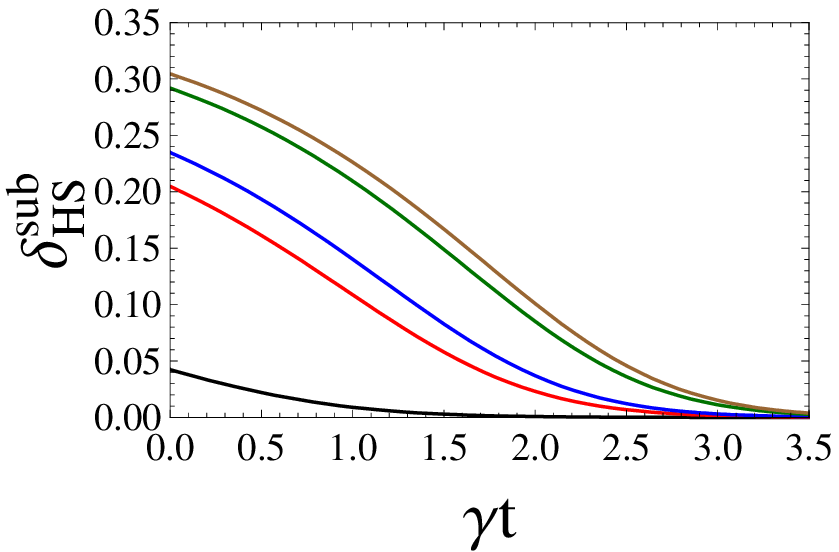}
\caption{Time evolution of non-Gaussianity  for $M$-photon-subtracted thermal states. We used the following parameters: $\bar n = 1.5$, $\bar n_R = 0.1$, $M = 1, 4, 5, 8, 9$. The upper plots correspond to the higher values of $M$.}
\label{fig-3}
\end{figure*}
We now take advantage of our previous results on the non-Gausssianity of damped PATSs \cite{MGM}. In Figure 4 we present a comparison between the time evolution of non-Gaussianity for photon-added 
(dotted curves) and photon-subtracted (continuous curves) corresponding to the same parameters of states and reservoir. 
\begin{figure*}[h]
\center
\includegraphics[width=5cm]{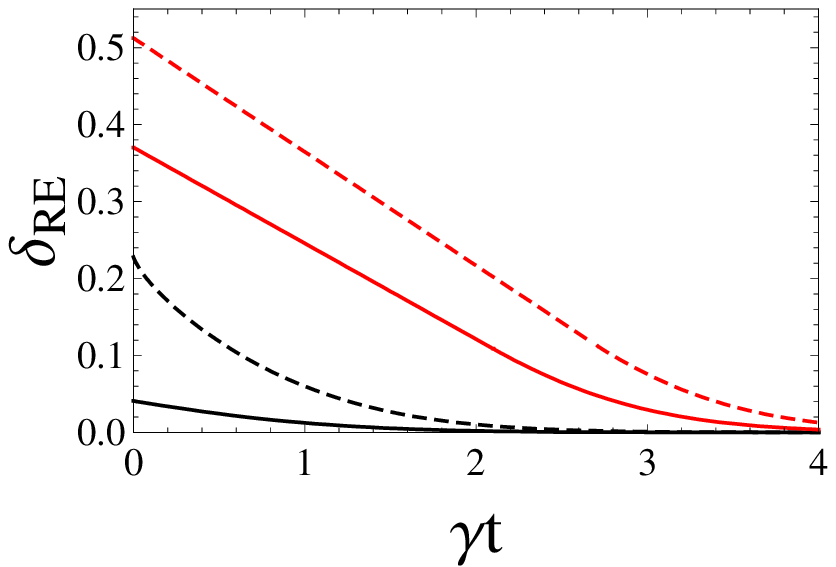}
\includegraphics[width=5cm]{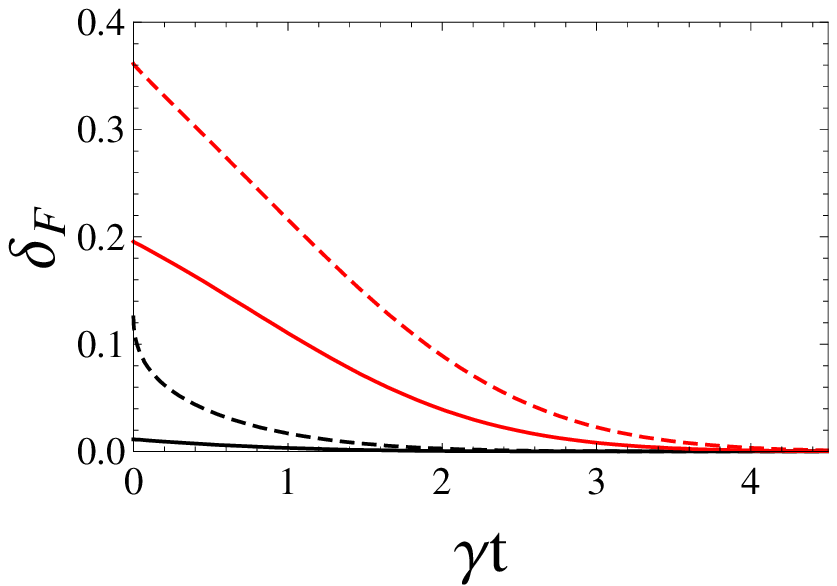}
\includegraphics[width=5cm]{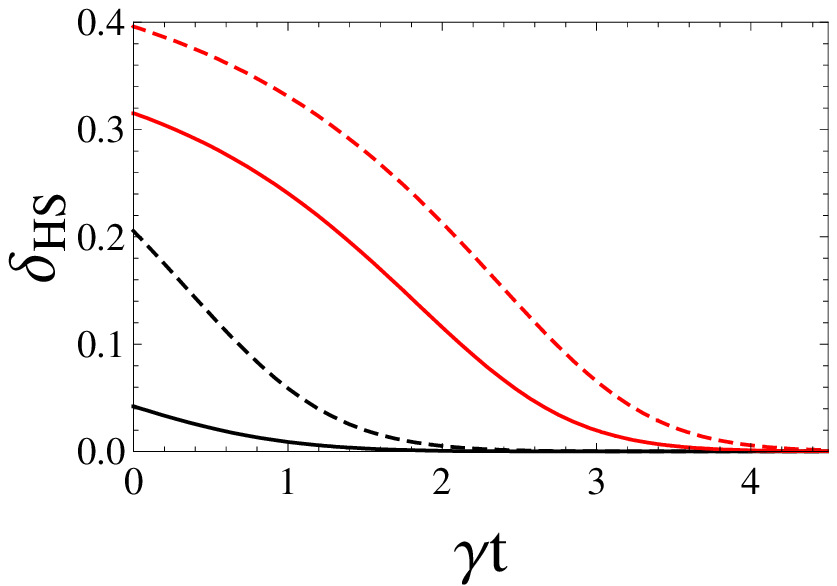}
\caption{Time evolution of non-Gaussianity  for $M$-photon-subtracted thermal states (continuous lines) and $M$-photon-added thermal states (dottted lines). We used the following parameters: $\bar n = 1.5$, $\bar n_R = 0.1$, $M$ = 1 (lower plots) and $M=10$ (upper plots).}
\label{fig-4}
\end{figure*}

\section{Conclusions}

Non-Gaussianity of pure states is often associated to their 
non-classicality. In this work we have examined the class of 
photon-subtracted one-mode thermal states, which are always mixed 
and classical. Subtraction of $M$ photons from a thermal state 
results in a photon-number distribution which turns out to be 
a negative binomial one whose stopping number is equal to $M+1$. 
The principal feature we have looked for was the non-Gaussianity 
of the PSTSs, as indicated by some recently introduced distance-type measures. Also investigated was the decrease of this property during the interaction of the field mode with a thermal reservoir described 
by the quantum optical master equation. We have shown that decaying 
of non-Gaussianity was consistently pointed out by three 
distance-type measures. Thus, the fidelity-based degree, 
the Hilbert-Schmidt one, and the entropic measure evolve monotonically, as expected for any good measure of non-Gaussianity \cite{GMM,MGM}. Moreover, we have compared this evolution to that 
of the photon-added thermal states  which, by contrast, are 
non-classical. We have found that a given PATS has a larger amount
of non-Gaussianity than the corresponding PSTS. This inequality between the degrees of non-Gaussianity of PATSs and PSTSs is maintained during their dissipative evolution. We conclude by stressing the significance of the agreement between the three measures of non-Gaussianity employed here, which is equally valid for both 
the PATSs and the PSTSs. Because the entropic measure 
$\delta_{\rm RE}[\hat \rho]$ is an exact one \cite{PT2013}, the other two, $\delta_{\rm F}[\hat \rho]$ and $\delta_{\rm HS}[\hat \rho]$, 
albeit approximate, are nevertheless reliable.

\ack{This work was supported by the Romanian National
Authority for Scientific Research through Grant No. PN-II-ID-
PCE-2011-3-1012 for the University of Bucharest.}

\appendix  
\section{Some useful formulae involving Gauss hypergeometric functions}

A Gauss hypergeometric function is the sum of the corresponding hypergeometric series:
\begin{equation}
_{2}F_{1}(a, b; c ; z):=\sum\limits_{n=0}^{\infty}\frac{(a)_{n}
(b)_{n}}{(c)_{n}}\frac{z^n}{n!}\,,\qquad (|z|<1), 
\label{a1}
\end{equation}
where $(a)_{n}:=\Gamma(a+n)/\Gamma(a)$ is Pochhammer's symbol standing for a rising factorial. This definition is extended by analytic continuation \cite{HTF}. Recall Pfaff's transformation formula
 \cite{HTF}, 
\begin{equation}
_{2}F_{1}(a, b\, ; c\, ; z)=(1-z)^{-b}\, {_{2}F_{1}}\left( c-a,\, 
b\, ; c\, ;\, \frac{z}{z-1}\right), 
\label{a2}
\end{equation}
as well as Murphy's expression of the Legendre polynomial of degree 
$l$ in terms of a Gauss hypergeometric function \cite{HTF}:
\begin{equation}
{\cal P}_l(z)={_{2}F_{1}}\left(-l, l+1; 1; \frac{1-z}{2} \right),
\qquad (l=0,1,2,3,...). 
\label{a3}
\end{equation}
The following sum \cite{HTF} has been used for obtaining the 
photon-number distribution\ (\ref{pndd}) in a damped PSTS: 
\begin{equation}
\sum_{n=0}^{\infty }\frac{(-\xi )_n}{n!}\, (-t)^n\; _{2}F_{1}(-n, b; c ; z) = (1+t)^\xi \; _{2}F_{1}\bigg( -\xi, b; c ; \frac{tz}{1+t}\bigg).
\label{a4}
\end{equation}

\end{document}